# BioDefect: The First Dataset for Defect Detection in Bioinformatics Software

Tianxiang Xu, Xiaoyan Zhu, Xin Lai, Xin Lian, Hangyu Cheng, Jiayin Wang†

School of Computer Science and Technology, Xi'an Jiaotong University

† Corresponding author. wangjiayin@mail.xjtu.edu.cn

## ABSTRACT

Software defect detection is a critical task in software engineering. However, no prior studies have specifically addressed defect detection in bioinformatics software. As a fundamental tool in modern biological research, bioinformatics software plays a crucial role in advancing life sciences. Unlike general-purpose software, bioinformatics software is typically developed by academic researchers, leading to significant differences in coding conventions, defect patterns, and other aspects compared to general software. These differences make existing defect detection datasets potentially unsuitable for bioinformatics software. Given that the performance of defect detection tasks is primarily influenced by both models and datasets, our experiments controlled for model-related factors and confirmed the limitations of existing datasets in bioinformatics software. To address this issue, we introduce BioDefect, the first dataset specifically designed for defect detection in bioinformatics software, aiming to overcome the limitations of existing datasets in this context. Unlike prior datasets, BioDefect includes complete source code repositories, preserving the actual contextual information of defective code, thereby more accurately reflecting real-world defect scenarios in bioinformatics software. Additionally, BioDefect mitigates issues related to label inconsistency and data leakage, ensuring high data quality and experimental reliability. To evaluate the effectiveness of BioDefect, we conduct a systematic assessment on nine language models (LMs), including DeepSeek-R1. The results demonstrate that BioDefect significantly enhances defect detection performance for bioinformatics software. Compared to existing datasets, BioDefect achieves an average F1-score improvement of 29.61% to 38.04% across all models, highlighting its superior advantages. Furthermore, we experiment with larger-scale models and more advanced techniques, but the improvements remain marginal, reinforcing the necessity of domain-specific datasets and models for defect detection in bioinformatics software. This study fills a critical research gap in bioinformatics software defect detection, laying a foundation for future studies in this field and offering new insights for improving bioinformatics software quality assurance.

## CCS CONCEPTS

• Security and privacy ~ Software security engineering

## KEYWORDS

Bioinformatics Software, Defect Detection, Language Model

## 1 INTRODUCTION

As an interdisciplinary field combining modern biology and information technology, bioinformatics has significantly advanced life sciences [1]. With the development of this field, bioinformatics software has emerged as a critical tool, playing an essential role in scientific research [2]. For instance, key tasks such as genomic sequence alignment, single-cell data analysis, and structural prediction heavily rely on high-quality bioinformatics software. However, unlike general-purpose software, bioinformatics software is typically developed by researchers rather than professional software engineers. As a result, it often lacks rigorous software development processes, leading to inconsistent code quality and a higher likelihood of defects. Such defects can introduce computational errors, data loss, or even unreliable research outcomes [3, 4]. Consequently, defect detection in bioinformatics software is crucial, particularly in applications related to human health and safety, where ensuring the reliability of bioinformatics software and minimizing potential risks is of utmost importance.

In recent years, language models (LMs) have achieved remarkable breakthroughs in natural language processing and software engineering, and they have been increasingly applied to software defect detection tasks [5-8]. Researchers have leveraged LMs to analyze source code and detect potential vulnerabilities or hidden functional defects [6]. Although LM-based methods have demonstrated strong performance on existing datasets, their applicability to bioinformatics software remains largely unverified. There are several challenges: (1) Bioinformatics software is primarily developed in Python and R, exhibiting diverse coding styles and development conventions, which may limit the direct applicability of existing defect detection methods. (2) Existing datasets are primarily designed for general software engineering projects, making them potentially insufficient for capturing defect patterns specific to bioinformatics software. Moreover, there are two key data quality issues in current datasets:

1. Label inconsistency, where the same sample is simultaneously labeled as "defective" and "non-defective".
2. Data leakage, where identical samples appear in both the training and test sets, leading to artificially inflated model performance.

These issues suggest that existing datasets may be unsuitable for bioinformatics software, making it difficult to effectively detect defects. To address these challenges, we introduce BioDefect, the first dataset specifically designed for defect detection in bioinformatics software. BioDefect is developed to enhance defect detection performance accurately and efficiently in this domain. Additionally, BioDefect consists of three independent bioinformatics software test sets, including software developed in Python and C/C++, ensuring comprehensive model evaluation across different bioinformatics software scenarios.

Compared to prior datasets [9], BioDefect is unique in that it contains entire source code repositories with defective code embedded in its actual development context, making it more representative of real-world bioinformatics software defects. Furthermore, BioDefect addresses the issue of label inconsistency through manual verification and a refined data collection strategy and mitigates data leakage by employing a time-series-based approach to split training and test sets (detailed in Section III).

To thoroughly evaluate the applicability of existing datasets and models to bioinformatics software, we conducted experiments using seven LMs of different scales, as well as two state-of-the-art open-source large language models (LLMs)—DeepSeek-R1 [10] and StarCoder2 [11]. First, we fine-tuned and evaluated these LMs on Devign [12] and ReVeal [9], where CodeT5 [13] and OPT [14] achieved the best performance, with F1-score of 58.11% and 57.61% on Devign, respectively. However, when evaluated on the three independent bioinformatics software test sets, model performance dropped significantly. For example, the F1-score of CodeT5 dropped by 49.41% to 55.15%, while OPT dropped by 47.22% to 54.83%. This drastic decline in performance was not an isolated case. LMs fine-tuned on ReVeal exhibited even greater performance degradation. These results indicate that existing datasets struggle to support defect detection in bioinformatics software, underscoring the necessity of developing dedicated datasets for this domain.

To validate the effectiveness of BioDefect, we fine-tuned the LMs on BioDefect and evaluated them on the three bioinformatics software test sets. The results show that LMs fine-tuned on BioDefect achieved superior performance compared to those fine-tuned on Devign and ReVeal. For example, on the Scanpy test set, LMs fine-tuned on BioDefect achieved F1-score improvements of 29.61% and 38.04% over Devign and ReVeal, respectively.

Additionally, we experimented with larger-scale models [10, 11] and more advanced techniques [15, 16] to further improve detection performance. However, the improvements were marginal, and these models still failed to completely distinguish defective code from non-defective code in bioinformatics software. This finding suggests that significant challenges remain in achieving accurate defect detection for bioinformatics software, potentially requiring fundamentally new approaches. Nonetheless, BioDefect, as the first dataset designed for defect detection in bioinformatics software, fills a critical research gap and establishes a foundation for future studies in this field.

In summary, the key contributions of this work are as follows:

- We propose the first defect detection task for bioinformatics software. To the best of our knowledge, this is the first study exploring defect detection in this domain, addressing a previously unstudied problem.
- We construct BioDefect, the first dataset specifically designed for defect detection in bioinformatics software, overcoming the limitations of existing datasets.
- We conduct a systematic evaluation of various LMs on bioinformatics software defect detection, revealing the limitations of existing datasets and highlighting their unsuitability for this domain. All code and datasets are publicly available at https://github.com/Ricardo1998-Xu/BioDefect

# 2 CHALLENGES

This section highlights the primary challenges faced by existing datasets in the task of defect detection for bioinformatics software.

## 2.1 Code Differences

The code in bioinformatics software differs significantly from general-purpose software in several key aspects:

**1) Programming Languages.** Due to its specialized application scenarios, bioinformatics software is predominantly developed in Python (for data processing and analysis) and R (for statistical analysis and visualization). However, most widely used defect detection datasets focus on C/C++ and Java [17-19]. This discrepancy in programming languages suggests that existing datasets may not be well-suited for detecting defects in bioinformatics software. Additionally, some bioinformatics software projects employ multi-language development, further complicating defect detection when using existing datasets.

**2) Coding Conventions.** Existing datasets are typically derived from large-scale, mature software engineering projects that adhere to standardized development processes, including consistent naming conventions, proper code documentation, and modular design [20, 21]. In contrast, bioinformatics software is often developed by academic researchers, many of whom lack formal software engineering backgrounds. As a result, bioinformatics software frequently exhibits diverse coding styles, non-standard practices, and a lack of clear module separation. These irregularities make it difficult for existing datasets to effectively address the defects present in such unstructured code.

**3) Defect Patterns.** Defects in bioinformatics software go beyond common code errors (such as syntax errors and logical flaws) and extend to algorithmic errors that can lead to incorrect biological inferences or inaccurate computational results. For example, an error in genomic sequence alignment might not cause a runtime failure or crash, yet it could yield misleading biological conclusions. Existing datasets primarily focus on conventional software defects (e.g., null pointer dereferences, memory leaks), which may be insufficient for detecting the unique types of defects encountered in bioinformatics software.

## 2.2 Label Inconsistency

Early datasets were manually constructed by injecting defects into otherwise correct code using predefined rules. However, these artificially introduced defects oversimplify real-world defect patterns and fail to capture their true complexity [22, 23]. To address this limitation, modern datasets [9, 12, 16, 17] primarily collect defect-fixing commits from real-world open-source software repositories, as illustrated in Figure 1.

Despite this improvement, such data collection strategies introduce label inconsistency issues. As shown in Figure 1, consider three defect-fixing commits at timestamps x, x+1, and x+2. Initially, in commit x, Function B contains a defect (later revised to Function B1 in commit x+1). Following the approach used in existing datasets such as ReVeal [9], Function B would be labeled as "defective", whereas other unchanged functions (Function A and C) would be labeled as "non-defective". However, as development progresses, commit x+1 introduces a fix for Function C, which is later updated to Function C1 in commit x+2. Under the same labeling strategy, Function C would now be labeled as "defective", while Function A and B1 remain "non-defective". This inconsistency results in two versions of Function C within the dataset, with one labeled as "defective" and the other as "non-defective". Some datasets [24] exacerbate this issue by automatically labeling all modified functions in a commit as "defective", leading to further inconsistencies.

Training models on such inconsistent labels introduces ambiguity, making defect detection in bioinformatics software significantly more challenging.

## 2.3 Data Leakage

Another major issue in existing datasets is data leakage, which has been widely recognized as a critical problem in defect detection research due to data duplication [25]. Data duplication can arise from several factors. For example, prior research [17] found that in the Devign [12], two commits corresponding to the same Common Vulnerabilities and Exposures (CVE) fix were sampled twice—one in the training set and the other in the test set.

Additionally, as illustrated in Figure 1, conventional data collection strategies result in multiple identical instances of Function A appearing in the dataset. When datasets are randomly split into training and test sets, these duplicated samples can be assigned to both sets, leading to unintentional data leakage. For instance, in the BigVul [24], up to 12.7% of samples are duplicated. Although some datasets attempt to deduplicate samples during collection, it is nearly impossible to eliminate all duplicates [26]. This data leakage artificially inflates the performance of existing defect detection models [27], but it does not contribute to real-world defect detection effectiveness.

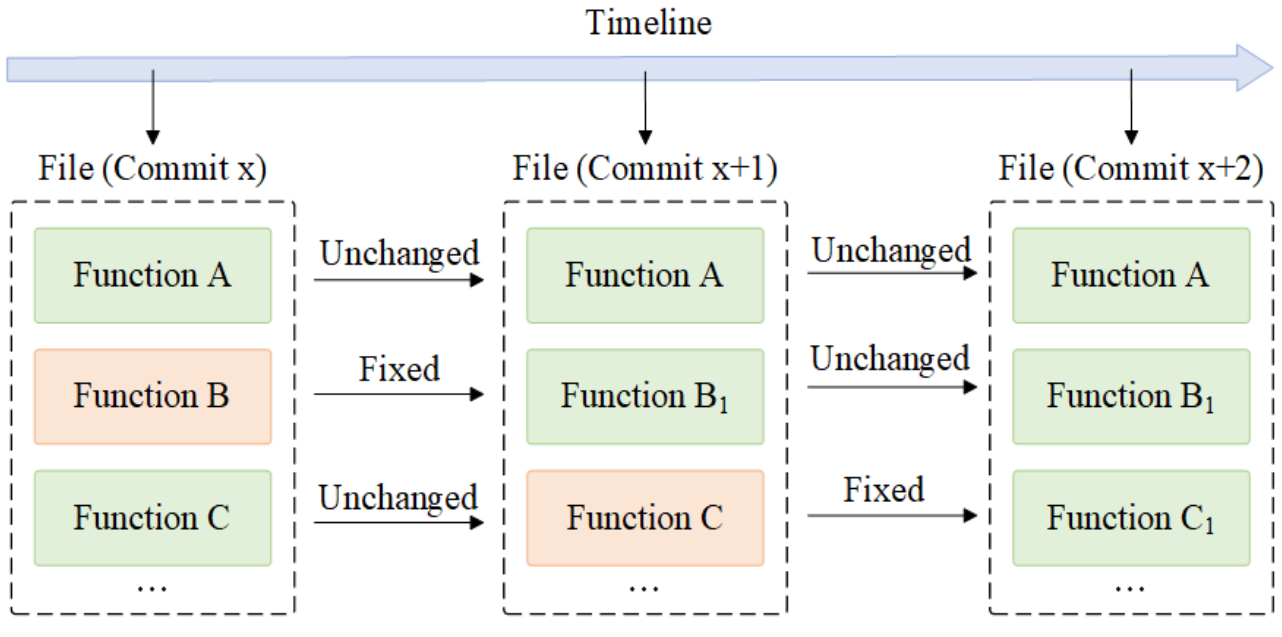

**Figure 1: Illustration of conventional dataset collection strategies. Green represents "non-defective" data, while orange represents "defective" data.**

# 3 BIODEFECT DATASET

In this section, we introduce BioDefect, the first dataset specifically designed for defect detection in bioinformatics software. BioDefect aims to address the limitations of existing datasets in this domain. Table 1 provides detailed information about BioDefect, including its primary training set and three independent test sets, ensuring comprehensive and precise model evaluation across different scenarios.

## 3.1 Project Selection

Given the differences between bioinformatics software and general-purpose software, the most effective approach for dataset construction is to base it on real-world bioinformatics software source code. GitHub hosts numerous open-source bioinformatics projects, some of which contain a substantial number of defect-related commits and corresponding fixes, making them suitable for dataset extraction. Specifically, we selected Scanpy [28], BWA [29], and Bowtie2 [30] to construct BioDefect based on the following criteria:

1. These software tools are widely used and well-maintained in the bioinformatics community, with extensive version histories.
2. They align with our defect detection requirements, covering diverse defect patterns that commonly appear in bioinformatics software.
3. They contain a large volume of publicly available commit histories, facilitating data collection and labeling.

**Table 1: BioDefect dataset details**

| Split | Project | Language | Project Size | # Defective Functions | # Non-Defective Functions |
|---|---|---|---|---|---|
| Train | BioDefect (0.0-1.9.0) | Python | 2054 | 1113 | 941 |
| Test | Scanpy (1.9.1) | Python | 941 | 63 | 878 |
| | BWA (0.7.17) | C/C++ | 286 | 10 | 276 |
| | Bowtie2 (2.5.1) | C/C++ | 648 | 14 | 634 |

### 3.2 Defect Data Collection

We chronologically sorted all commits from the selected bioinformatics software projects. Two experts in software engineering reviewed each commit to determine whether it was related to a defect fix. If a commit involved function-level defect fixes, we extracted the pre-fix version of the function and labeled it as "defective". Other unchanged functions were left unprocessed to prevent incorrect labeling. For example, in Figure 1, we extract Function B from commit x and Function C from commit x+1, labeling them as "defective". Other functions remain unlabeled. This strategy prevents inconsistencies and duplication in BioDefect, distinguishing it from existing datasets.

To ensure alignment with real-world scenarios, models should be trained on historical data and used to predict future defects. Therefore, we adopted a time-series-based data splitting strategy when dividing BioDefect into training and test sets. Specifically, we sorted all collected defect data based on software version history:

1. 85% of the earliest versions were allocated to the training set.
2. The remaining 15% of later versions were designated as test sets.

For instance, in the Scanpy project, the training set includes versions 0.0 to 1.9.0, covering commits from January 29, 2017, to April 1, 2022. The test set includes versions 1.9.1 to 1.10.1, covering commits from August 12, 2022, to April 9, 2024. Notably, these dates correspond to Scanpy's defect-related commits, and all dataset details are publicly available.

### 3.3 Non-Defect Data Collection

To construct the non-defect samples in BioDefect, we extracted functions from two specific software versions—one for the training set and another for the test set. For example, in the Scanpy project, we extracted all functions from version 1.9.0, labeling them as "non-defective" to build the training set's non-defect data. This approach reflects real-world conditions, where developers cannot predict in advance which functions will require fixes in the future. For the test set, we extracted all functions from version 1.9.1. If a function was subsequently fixed in versions 1.9.1 to 1.10.1, we labeled it as "defective"; otherwise, it remained "non-defective". The same data collection strategy was applied to BWA and Bowtie2.

**Table 2: Summary statistics of the BioDefect dataset. CC: Cyclomatic complexity, HC: Halstead complexity. '25%' and '75%' represents 25th quantile and 75th quantile, respectively. All the statistics are on function granularity.**

| Statistic | Mean | Std | Min | 25% | Median | 75% | Max |
|---|---|---|---|---|---|---|---|
| LOC | 70 | 78 | 2 | 14 | 40 | 102 | 401 |
| Tokens | 480 | 585 | 12 | 114 | 264 | 616 | 3466 |
| CC | 14 | 18 | 1 | 2 | 6 | 16 | 108 |
| HC | 268 | 510 | 0 | 5 | 62 | 255 | 3349 |

In contrast to previous datasets [9, 12, 17], which sometimes trained models using future data to evaluate past defects, BioDefect adopts a fully independent test set covering the entire source code repository (including both defective and non-defective functions). This approach ensuring a realistic representation of defect scenarios in bioinformatics software. Furthermore, BioDefect adopts a strict time-series partitioning strategy, eliminating any overlap between the training and test sets to prevent data leakage. This strategy enhances the accuracy and practical relevance of model evaluations, making the results more reliable for real-world applications.

### 3.4 Analysis of BioDefect Statistics

We conducted a statistical analysis of BioDefect using lines of code (LOC), token count, Cyclomatic complexity and Halstead complexity to further understand the complexity and diversity of bioinformatics software code, as summarized in Table 2. BioDefect exhibits a wide range of function sizes, encompassing simple, short functions as well as highly complex ones. The high standard deviation indicates significant variation, which aligns with the inherent diversity of bioinformatics software. Additionally, we observed outliers in both Cyclomatic complexity and Halstead complexity, suggesting that existing datasets might struggle with these highly complex functions.

Furthermore, we analyzed the distribution of defective and non-defective functions in BioDefect based on LOC (with similar trends observed for token count, Cyclomatic complexity and Halstead complexity), as shown in Figure 2. BioDefect primarily consists of short functions, mostly under 50 LOC. However, as LOC increases, the proportion of defective functions also rises. For instance, in functions ranging 10-20 LOC, defective functions account for approximately 31%, whereas in functions ranging 40-80 LOC, this proportion rises to about 72%. Notably, for functions exceeding 80 LOC, the proportion of defective functions surpasses 83%. This imbalance between LOC and defect proportion further highlights the challenges in bioinformatics software defect detection.

## 4 EMPIRICAL STUDY

To evaluate the effectiveness of existing datasets and BioDefect in bioinformatics software defect detection, we formulate the following two core research questions:

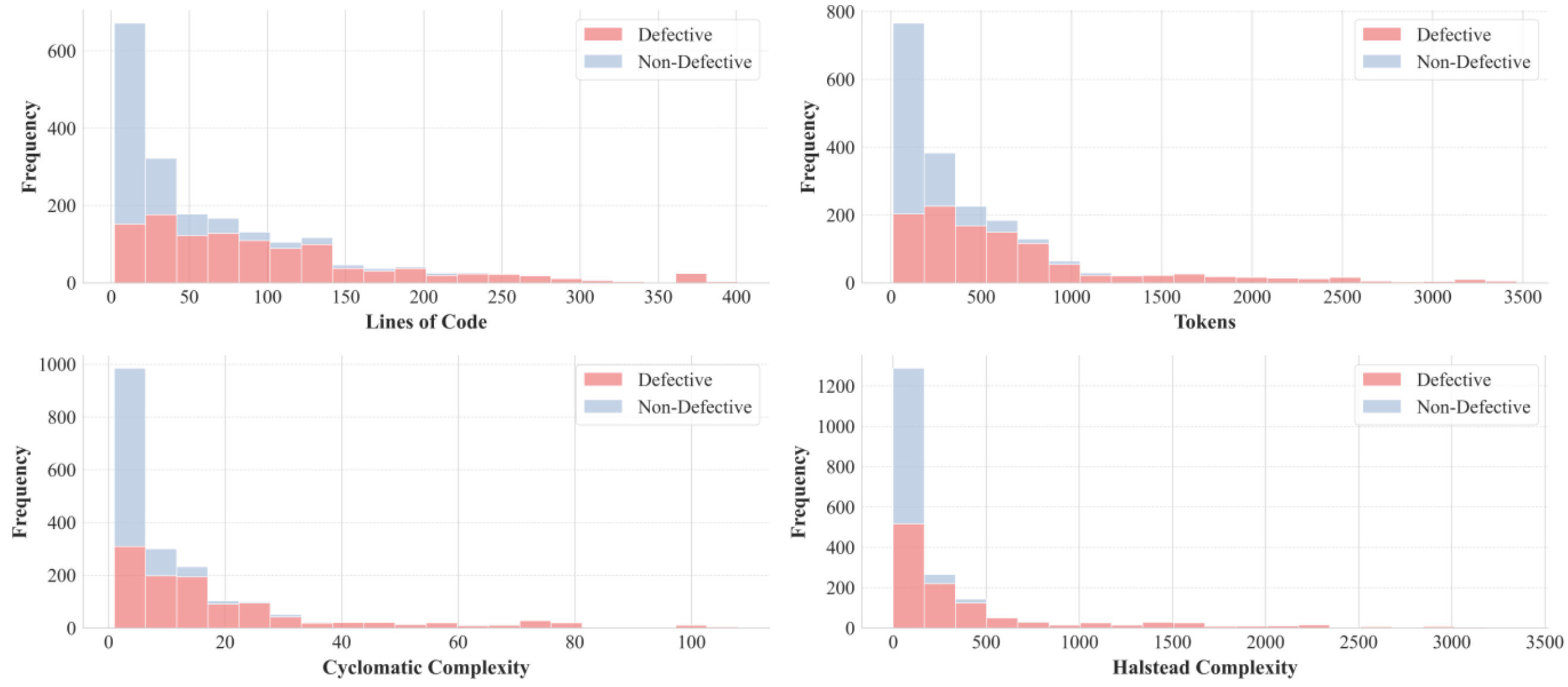

**Figure 2: Distribution of defective and non-defective functions in BioDefect**

- RQ1: Are existing datasets suitable for bioinformatics software defect detection?
- RQ2: How effective is BioDefect in the bioinformatics software defect detection?

## 4.1 Study Subject

**Datasets**. For RQ1, we aim to verify the applicability of existing datasets to bioinformatics software, selecting Devign [12] and ReVeal [9] as representative datasets for our experiments. These two datasets have been widely adopted in defect detection research [6-8, 13, 31-34], but their suitability for bioinformatics software has not been thoroughly examined. By comparing them with BioDefect, we reveal the limitations of existing datasets in bioinformatics software defect detection. For RQ2, we primarily utilize BioDefect to comprehensively assess its effectiveness in this task.

**Models.** As shown in Table 3, we select nine LMs of varying sizes to analyze their performance in bioinformatics software defect detection. Specifically, we fine-tune seven LMs with fewer than 1 billion parameters to evaluate both the applicability of existing datasets and the effectiveness of BioDefect. To eliminate potential performance degradation caused by model capacity limitations, we further fine-tune two LLMs (DeepSeek-R1 [10] and StarCoder2 [11]), each exceeding 1 billion parameters. Notably, in the advanced experiments of RQ2, we choose OPT as the primary experimental model.

**Experimental Settings.** We implement our fine-tuning framework based on existing benchmarks [17, 31] and conduct all experiments accordingly. All models are optimized using the AdamW algorithm, fine-tuned for 10 epochs with Cross-Entropy Loss and Focal Loss [15]. For models with fewer than 1B parameters, we adopt a learning rate of 2e-5, whereas for larger models, we set the learning rate to 1e-6. The pre-trained weights for all models are sourced from Hugging Face Models [39]. All fine-tuning and evaluation tasks are conducted on two NVIDIA GeForce RTX 4090 GPUs. To ensure the reliability of our results, all reported outcomes are averaged over three random seeds.

**Evaluation Metrics.** To accurately assess the performance of LMs in bioinformatics software defect detection, we employ multiple evaluation metrics. In addition to the commonly used precision, recall, and F1-score, we incorporate Balance and Matthews Correlation Coefficient (MCC), as these metrics comprehensively evaluate model performance across both positive and negative classes, mitigating biases caused by class imbalance. These metrics are widely used in related research [40, 41]. Notably, the range for Balance is [0, 1], while MCC ranges from [-1, 1].

**Table 3: Information on LMs of different scales**

| Model | Parameters | Architecture | Methods |
|---|---|---|---|
| BERT [35] | 110M | Encoder | Fine-tuning |
| CodeBERT [36] | 125M | Encoder | Fine-tuning |
| T5 [37] | 220M | Encoder-Decoder | Fine-tuning |
| CodeT5 [13] | 220M | Encoder-Decoder | Fine-tuning |
| CodeT5+ [8] | 770M | Encoder-Decoder | Fine-tuning |
| OPT [14] | 125M | Decoder | Fine-tuning |
| CodeGen [38] | 350M | Decoder | Fine-tuning |
| DeepSeek-R1 [10] | 1.5B | Decoder | Fine-tuning Checkpoint |
| StarCoder2 [11] | 3B | Decoder | Fine-tuning Checkpoint |

## 4.2 RQ1: Are Existing Datasets Suitable for Bioinformatics Software Defect Detection?

To address this research question, we fine-tune nine LMs on existing defect detection datasets and evaluate their suitability for bioinformatics software defect detection. Specifically, we first fine-tune seven LMs with fewer than 1 billion parameters on the Devign and ReVeal datasets, then evaluate them on both their respective test sets and the Scanpy test set. The results are presented in Table 4. In this table, "Train=Devign" and "Test=Devign" indicate that the model was fine-tuned and evaluated on Devign, while "Train=Devign" and "Test=Scanpy" signify that the model was fine-tuned on Devign but evaluated on Scanpy. Note that "/" indicates cases where the denominator is zero, making the computation of certain metrics meaningless.

**Results.** Existing defect detection datasets are unsuitable for bioinformatics software and fail to effectively identify defects in such applications. A striking revelation from our study is that the performance of models fine-tuned on these datasets is significantly overestimated, as they struggle to learn the complex defect patterns specific to bioinformatics software. For example, on the Devign dataset, CodeT5 achieves an F1-score of 0.5811.

**Table 4: Performance of fine-tuned LMs on bioinformatics software using existing datasets**

| Model | Train | Test | Precision | Recall | F1 | Balance | MCC |
|---|---|---|---|---|---|---|---|
| BERT | Devign | Devign | 0.5913 | 0.5599 | 0.5741 | 0.6096 | 0.2324 |
| | | Scanpy | 0.0661 | 0.4974 | 0.1161 | 0.4842 | -0.0033 |
| | ReVeal | ReVeal | 0.4562 | 0.2631 | 0.3221 | 0.4782 | 0.2893 |
| | | Scanpy | 0 | 0 | / | 0.2929 | -0.0106 |
| CodeBERT | Devign | Devign | 0.6097 | 0.5267 | 0.5639 | 0.6069 | 0.2445 |
| | | Scanpy | 0.0542 | 0.4232 | 0.0954 | 0.4392 | -0.0434 |
| | ReVeal | ReVeal | 0.6260 | 0.3079 | 0.4095 | 0.5103 | 0.3972 |
| | | Scanpy | 0.0256 | 0.0053 | 0.0088 | 0.2966 | 0.0047 |
| T5 | Devign | Devign | 0.5635 | 0.4574 | 0.5032 | 0.5593 | 0.1613 |
| | | Scanpy | 0.0373 | 0.3122 | 0.0664 | 0.3736 | -0.1102 |
| | ReVeal | ReVeal | 0.4533 | 0.2855 | 0.3496 | 0.4941 | 0.3062 |
| | | 0.9274 | 0.0053 | 0.9935 | 0.0417 | -0.0074 | 0.5004 |
| CodeT5 | Devign | Devign | 0.6250 | 0.5493 | 0.5811 | 0.6184 | 0.2710 |
| | | Scanpy | 0.0484 | 0.4551 | 0.0870 | 0.3914 | -0.0995 |
| | ReVeal | ReVeal | 0.4855 | 0.3901 | 0.4283 | 0.5673 | 0.3780 |
| | | Scanpy | 0 | 0 | / | 0.2929 | -0.0087 |
| CodeT5+ | Devign | Devign | 0.5767 | 0.4659 | 0.5122 | 0.5647 | 0.1771 |
| | | Scanpy | 0.0441 | 0.2593 | 0.0746 | 0.3954 | -0.0745 |
| | ReVeal | ReVeal | 0.5259 | 0.2467 | 0.3358 | 0.4670 | 0.3158 |
| | | Scanpy | 0 | 0 | / | 0.2929 | -0.0119 |
| OPT | Devign | Devign | 0.5974 | 0.5572 | 0.5761 | 0.6131 | 0.2397 |
| | | Scanpy | 0.0589 | 0.4815 | 0.1039 | 0.4330 | -0.0644 |
| | ReVeal | ReVeal | 0.4564 | 0.3557 | 0.3956 | 0.5430 | 0.3439 |
| | | Scanpy | 0.1667 | 0.0053 | 0.0103 | 0.2966 | 0.0357 |
| CodeGen | Devign | Devign | 0.6176 | 0.5153 | 0.5609 | 0.6058 | 0.2497 |
| | | Scanpy | 0.0670 | 0.8783 | 0.1245 | 0.3704 | 0.0060 |
| | ReVeal | ReVeal | 0.3979 | 0.2974 | 0.3352 | 0.5017 | 0.2795 |
| | | Scanpy | 0.0417 | 0.0106 | 0.0169 | 0.3003 | -0.0012 |

However, when evaluated on Scanpy, its F1-score drops to 0.087, representing a dramatic decline of 49.41%. In more extreme cases, certain LMs fine-tuned on the ReVeal dataset yield a Precision of 0 when tested on bioinformatics software. Importantly, this sharp performance decline is not an isolated case but a consistent trend across all LMs and evaluation metrics.

However, as Devign [12] and ReVeal [9] primarily consist of C/C++ code, we conduct additional experiments on two bioinformatics software projects written in C/C++ (BWA and Bowtie2) to rule out the possibility that the performance degradation is merely due to differences in programming languages. The results, presented in Table 5, further confirm the limitations of existing datasets. Performance on BWA and Bowtie2 remains significantly lower than expected, with some LMs fine-tuned on ReVeal failing to produce any correct predictions—classifying all samples as non-defective. Such a failure is entirely unacceptable in real-world applications, underscoring the unsuitability of existing datasets for bioinformatics software defect detection.

Since defect detection performance may also be influenced by model scale, we further fine-tune two LLMs—DeepSeek-R1 [10] and StarCoder2 [11]—to investigate whether LLMs can compensate for the limitations of existing datasets. As shown in Table 6, although these LLMs outperform smaller ones on the three bioinformatics software test sets, they still exhibit performance degradation when compared to results on the Devign and ReVeal test sets. Moreover, comparing the performance of the seven LMs with the two LLMs reveals an important trend: LLMs exhibit stronger generalization capabilities, relying less on specific training data, and demonstrate more stable performance across the three bioinformatics software test sets. Additionally, they are less affected by programming language differences. However, even LLMs fail to overcome the fundamental limitations of existing datasets, indicating that dataset suitability is the critical factor affecting bioinformatics software defect detection, rather than merely model capacity constraints.

**Table 5: Evaluation results on C/C++ bioinformatics software**

| Model | Train | Test | Precision | Recall | F1 | Balance | MCC |
|---|---|---|---|---|---|---|---|
| BERT | Devign | Bowtie2 | 0.0123 | 0.3333 | 0.0237 | 0.3674 | -0.0722 |
| | | BWA | 0.0132 | 0.2667 | 0.0251 | 0.2926 | -0.1499 |
| | ReVeal | Bowtie2 | 0 | 0 | / | 0.2929 | / |
| | | BWA | 0 | 0 | / | 0.2929 | -0.0137 |
| CodeBERT | Devign | Bowtie2 | 0.0204 | 0.5476 | 0.0393 | 0.4731 | -0.0083 |
| | | BWA | 0.0296 | 0.5667 | 0.0562 | 0.4176 | -0.0379 |
| | ReVeal | Bowtie2 | 0 | 0 | / | 0.2929 | / |
| | | BWA | 0 | 0 | / | 0.2929 | -0.0113 |
| T5 | Devign | Bowtie2 | 0.0194 | 0.3809 | 0.0369 | 0.4635 | -0.0113 |
| | | BWA | 0.0157 | 0.2333 | 0.0294 | 0.3409 | -0.1082 |
| | ReVeal | Bowtie2 | 0 | 0 | / | 0.2929 | / |
| | | BWA | 0 | 0 | / | 0.2929 | -0.0113 |
| CodeT5 | Devign | Bowtie2 | 0.0154 | 0.3571 | 0.0296 | 0.4208 | -0.0420 |
| | | BWA | 0.0209 | 0.3333 | 0.0394 | 0.3807 | -0.0856 |
| | ReVeal | Bowtie2 | 0.6667 | 0.0476 | 0.0889 | 0.3266 | 0.1736 |
| | | BWA | 0 | 0 | / | 0.2929 | -0.0160 |
| CodeT5+ | Devign | Bowtie2 | 0.0140 | 0.3809 | 0.0270 | 0.3858 | -0.0673 |
| | | BWA | 0.0322 | 0.5333 | 0.0606 | 0.4667 | -0.0115 |
| | ReVeal | Bowtie2 | 0 | 0 | / | 0.2929 | -0.0071 |
| | | BWA | 0 | 0 | / | 0.2929 | -0.0113 |
| OPT | Devign | Bowtie2 | 0.0166 | 0.5238 | 0.0322 | 0.4058 | -0.0427 |
| | | BWA | 0.0147 | 0.2667 | 0.0278 | 0.3053 | -0.1430 |
| | ReVeal | Bowtie2 | 0 | 0 | / | 0.2929 | -0.0058 |
| | | BWA | 0 | 0 | / | 0.2929 | / |
| CodeGen | Devign | Bowtie2 | 0.0208 | 0.9524 | 0.0407 | 0.2986 | -0.0422 |
| | | BWA | 0.0340 | 0.9667 | 0.0657 | 0.2960 | -0.0411 |
| | ReVeal | Bowtie2 | 0 | 0 | / | 0.2929 | -0.0123 |
| | | BWA | 0 | 0 | / | 0.2928 | -0.0228 |

**Table 6: Performance of fine-tuned LLMs on bioinformatics software using existing datasets**

| Method | Train | Test | Precision | Recall | F1 | Balance | MCC |
|---|---|---|---|---|---|---|---|
| DeepSeek-R1 | Devign | Devign | 0.5180 | 0.2454 | 0.3328 | 0.4490 | 0.0622 |
| | | Scanpy | 0.1922 | 0.3757 | 0.2537 | 0.5511 | 0.1941 |
| | | Bowtie2 | 0.1235 | 0.1905 | 0.1492 | 0.4272 | 0.1302 |
| | | BWA | 0.1133 | 0.4000 | 0.1766 | 0.5682 | 0.1600 |
| | ReVeal | ReVeal | 0.0888 | 0.3976 | 0.1196 | 0.3965 | -0.0395 |
| | | Scanpy | 0.1329 | 0.2857 | 0.1429 | 0.3886 | 0.0174 |
| | | Bowtie2 | 0.1094 | 0.6191 | 0.0911 | 0.3793 | 0.0593 |
| | | BWA | 0.0182 | 0.4333 | 0.0349 | 0.3300 | -0.0988 |
| StarCoder2 | Devign | Devign | 0.5200 | 0.1341 | 0.2095 | 0.3822 | 0.0417 |
| | | Scanpy | 0.1915 | 0.3386 | 0.2446 | 0.5267 | 0.1823 |
| | | Bowtie2 | 0.1576 | 0.2381 | 0.1895 | 0.4609 | 0.1717 |
| | | BWA | 0.1267 | 0.3000 | 0.1751 | 0.5013 | 0.1462 |
| | ReVeal | ReVeal | 0.2429 | 0.0717 | 0.1028 | 0.3432 | 0.0742 |
| | | Scanpy | 0.1908 | 0.2434 | 0.2131 | 0.4623 | 0.1502 |
| | | Bowtie2 | 0.200 | 0.1429 | 0.1667 | 0.3939 | 0.1570 |
| | | BWA | 0.0303 | 0.0667 | 0.0417 | 0.3387 | 0.0039 |

**Findings-RQ1:** Existing datasets are unsuitable for bioinformatics software defect detection and exhibit significant limitations.

## 4.3 RQ2: How Effective is BioDefect in Bioinformatics Software Defect Detection?

Based on the experimental results of RQ1, we found that existing defect detection datasets are not suitable for bioinformatics software, highlighting the importance of constructing datasets that better align with the characteristics of bioinformatics software. Therefore, in this RQ, we further verify the effectiveness of BioDefect in the defect detection task for bioinformatics software. First, we fine-tune seven LMs using BioDefect and evaluate their performance on three bioinformatics software test sets, with the results presented in Table 7.

**Table 7: Performance of fine-tuned LMs on bioinformatics software using BioDefect**

| Model | Train | Test | Precision | Recall | F1 | Balance | MCC |
|---|---|---|---|---|---|---|---|
| BERT | BioDefect | Scanpy | 0.3636 | 0.5344 | 0.4297 | 0.6670 | 0.3910 |
| | | Bowtie2 | 0.0734 | 0.3333 | 0.1182 | 0.5229 | 0.1240 |
| | | BWA | 0.0742 | 0.4000 | 0.1247 | 0.5571 | 0.1269 |
| CodeBERT | BioDefect | Scanpy | 0.4107 | 0.4921 | 0.4459 | 0.6389 | 0.4056 |
| | | Bowtie2 | 0 | 0 | / | 0.2929 | -0.0138 |
| | | BWA | 0.0667 | 0.0667 | 0.0667 | 0.3395 | 0.0333 |
| T5 | BioDefect | Scanpy | 0.3186 | 0.4550 | 0.3728 | 0.6112 | 0.3262 |
| | | Bowtie2 | 0.0231 | 0.1191 | 0.0373 | 0.3651 | -0.0088 |
| | | BWA | 0.0597 | 0.7333 | 0.1100 | 0.6163 | 0.1163 |
| CodeT5 | BioDefect | Scanpy | 0.3958 | 0.4974 | 0.4289 | 0.6405 | 0.3898 |
| | | Bowtie2 | 0.1964 | 0.2381 | 0.2020 | 0.4607 | 0.1880 |
| | | BWA | 0.1279 | 0.3667 | 0.1868 | 0.5460 | 0.1648 |
| CodeT5+ | BioDefect | Scanpy | 0.4929 | 0.3863 | 0.4292 | 0.5655 | 0.3990 |
| | | Bowtie2 | 0.0666 | 0.4762 | 0.1142 | 0.6078 | 0.1267 |
| | | BWA | 0.0904 | 0.8333 | 0.1617 | 0.7323 | 0.2012 |
| OPT | BioDefect | Scanpy | 0.4440 | 0.4973 | 0.4659 | 0.6429 | 0.4275 |
| | | Bowtie2 | 0.0667 | 0.0714 | 0.0690 | 0.3434 | 0.0557 |
| | | BWA | 0.0815 | 0.2333 | 0.1078 | 0.4518 | 0.0876 |
| CodeGen | BioDefect | Scanpy | 0.4284 | 0.1058 | 0.1682 | 0.3676 | 0.1855 |
| | | Bowtie2 | 0.0632 | 0.4524 | 0.1086 | 0.5862 | 0.1161 |
| | | BWA | 0.0690 | 0.6000 | 0.1193 | 0.5514 | 0.1177 |

**Results.** BioDefect significantly improves the performance of defect detection in bioinformatics software. For instance, in the Scanpy test set, the F1-score of the OPT model fine-tuned with Devign and ReVeal were 0.1039 and 0.0103, respectively (see Table 4). However, after fine-tuned with BioDefect, the F1-score increased to 0.4659 (see Table 7), representing improvements of 36.2% and 45.56% compared to Devign and ReVeal, respectively. Notably, this performance improvement is not limited to the OPT model but is observed across all LMs. Specifically, in the evaluation on Scanpy, the average F1-score of the seven LMs fine-tuned with Devign was 0.0954, while that of those fine-tuned with ReVeal was 0.0111. In contrast, the average F1-score of the seven LMs fine-tuned with BioDefect reached 0.3915, representing improvements of 29.61% and 38.04%, respectively. For the MCC metric, LMs fine-tuned with BioDefect achieved an average MCC of 0.3607, improving by 41.63% and 36.08%, respectively.

Further analysis of performance on the BWA and Bowtie2 test sets shows that BioDefect continues to enhance defect detection effectiveness. For example, in the evaluation of Bowtie2, the average F1-score of LMs fine-tuned with Devign was 0.0328 (see Table 5), whereas LMs fine-tuned with BioDefect achieved an average F1-score of 0.1082 (see Table 7), representing a 7.54% improvement. The MCC metric increased from -0.0409 (Devign-based) and 0.0371 (ReVeal-based) to 0.084, improving by 12.49% and 4.69%, respectively. In the evaluation of BWA, the average MCC of LMs fine-tuned with Devign was -0.0825, while that of LMs fine-tuned with ReVeal was -0.0144. In contrast, the average

MCC of LMs fine-tuned with BioDefect was 0.1211, representing improvements of 20.36% and 13.55%, respectively.

Although the improvement observed in BWA and Bowtie2 is less pronounced than in Scanpy, we attribute this to differences in programming languages. BWA and Bowtie2 are developed in C/C++, whereas BioDefect is primarily constructed based on Python code. However, despite the language difference, BioDefect still effectively enhances defect detection performance. In contrast, Devign and ReVeal, which are composed of C/C++ code, performed poorly, often predicting all samples as the same class. This further suggests that defect patterns in bioinformatics software significantly differ from those in existing datasets, to the extent that this difference outweighs the impact of programming language alone.

Nevertheless, while BioDefect significantly improves defect detection performance in bioinformatics software, the performance is still insufficient for real-world applications. Therefore, we further explore the impact of larger-scale models and more advanced techniques to enhance performance.

1) Larger-Scale Models: We experimented with DeepSeek-R1 [10] and StarCoder2 [11], with results shown in Table 8. The results indicate that larger-scale models fail to improve defect detection performance for bioinformatics software. Typically, a larger parameter size corresponds to stronger capabilities, but this trend does not hold in the defect detection task for bioinformatics software. For example, in the evaluation of Scanpy, OPT (F1-score: 0.4659, MCC: 0.4275) and CodeBERT (F1-score: 0.4459, MCC: 0.4056) performed best (see Table 7), while StarCoder2 (F1-score: 0.1436, MCC: 0.0419) and DeepSeek-R1 (F1-score: 0.3134, MCC: 0.2886) lagged significantly behind (see Table 8).

We believe this is because LLMs are primarily designed for code completion and generation tasks, focusing more on overall semantic information, whereas defect detection requires fine-grained recognition of erroneous patterns. Consequently, increasing the parameter size does not yield the expected performance gain. However, we observed a phenomenon similar to RQ1, where LLMs exhibit stronger generalization capabilities and perform more stably in cross-language evaluations. This may be attributed to their superior contextual understanding, enabling them to capture commonalities in code patterns across different languages. This also explains why LLMs perform more stably in cross-language bioinformatics software but do not offer advantages in single-language tasks.

**Table 8: Performance of fine-tuned LLMs on bioinformatics software using BioDefect**

| Model | Train | Test | Precision | Recall | F1 | Balance | MCC |
|---|---|---|---|---|---|---|---|
| DeepSeek-R1 | BioDefect | Scanpy | 0.2064 | 0.6508 | 0.3134 | 0.7223 | 0.2886 |
| | | Bowtie2 | 0.2368 | 0.6429 | 0.3461 | 0.7454 | 0.3695 |
| | | BWA | 0.0926 | 0.5000 | 0.1563 | 0.6248 | 0.1514 |
| StarCoder2 | BioDefect | Scanpy | 0.0841 | 0.5714 | 0.1436 | 0.5173 | 0.0419 |
| | | Bowtie2 | 0.0175 | 0.6428 | 0.0340 | 0.3711 | -0.0584 |
| | | BWA | 0.0256 | 0.5667 | 0.0489 | 0.3598 | -0.1086 |

2) More Advanced Techniques: Given that larger models did not achieve the desired results, we conducted an in-depth analysis of challenging samples and monitored model behavior during evaluation to investigate the reasons behind incorrect predictions. Analysis revealed that due to the complex structure of bioinformatics software code, non-defective code is more stable and easier to learn during fine-tuning, whereas defective code exhibits sparser and more complex patterns, making gradient updates difficult to converge. As a result, models tend to predict non-defective categories, leading to many defect samples being misclassified as non-defective. Based on an extensive literature review, we experimented with Class Weights [16] and Focal Loss [15], two techniques that help models focus on critical or hard-to-detect samples (e.g., defective code) and have demonstrated effectiveness in binary classification tasks [17, 42]. We integrated Class Weights and Focal Loss into the fine-tuning framework, selecting OPT for fine-tuning with BioDefect and evaluating on the Scanpy test set, with results shown in Tables 9 and 10.

Class Weights adjust the loss calculation by assigning higher weights to defect class data, making the model more attentive to defective code during fine-tuning. To determine the optimal weight ratio, we experimented with different settings. Notably, when Weight=1, it corresponds to standard Cross-Entropy Loss. Similar to results in [16, 17], applying Class Weights improved F1-score and MCC metrics. For instance, when the defect class weight was ten times that of the non-defect class, the highest values were achieved (F1-score: 0.5133, MCC: 0.4784), improving by 4.74% and 5.09%, respectively.

Focal Loss introduces a parameter $\gamma$ to reduce the weight of easily classified samples, encouraging the model to focus more on hard-to-classify defective code. Notably, when $\gamma$=0, it corresponds to standard Cross-Entropy Loss. When $\gamma$=5, F1-score was0.4822, and MCC was 0.4478, representing slight improvements of 1.63% and 2.03%, respectively. However, the limited improvement further illustrates the challenges of bioinformatics software defect detection, suggesting the need for novel approaches.

**Table 9: Impact of Class Weights on bioinformatics software defect detection**

| Weight | Precision | Recall | F1 | Balance | MCC |
|---|---|---|---|---|---|
| 1 | 0.4440 | 0.4973 | 0.4659 | 0.6429 | 0.4275 |
| 2 | 0.4141 | 0.6032 | 0.4837 | 0.7144 | 0.4521 |
| 5 | 0.3997 | 0.5873 | 0.4610 | 0.7007 | 0.4297 |
| 10 | 0.4662 | 0.5767 | 0.5133 | 0.6986 | 0.4784 |
| 20 | 0.4317 | 0.5873 | 0.4791 | 0.7012 | 0.4497 |

**Table 10: Impact of Focal Loss on bioinformatics software defect detection**

| $\gamma$ | Precision | Recall | F1 | Balance | MCC |
|---|---|---|---|---|---|
| 0 | 0.4440 | 0.4973 | 0.4659 | 0.6429 | 0.4275 |
| 0.2 | 0.3335 | 0.5450 | 0.3922 | 0.6673 | 0.3586 |
| 0.5 | 0.4090 | 0.5079 | 0.4464 | 0.6493 | 0.4080 |
| 1 | 0.3987 | 0.4498 | 0.4217 | 0.6093 | 0.3789 |
| 2 | 0.3957 | 0.4709 | 0.4250 | 0.6236 | 0.3828 |
| 5 | 0.5032 | 0.4656 | 0.4822 | 0.6213 | 0.4478 |
| 10 | 0.3711 | 0.4074 | 0.3880 | 0.5795 | 0.3423 |

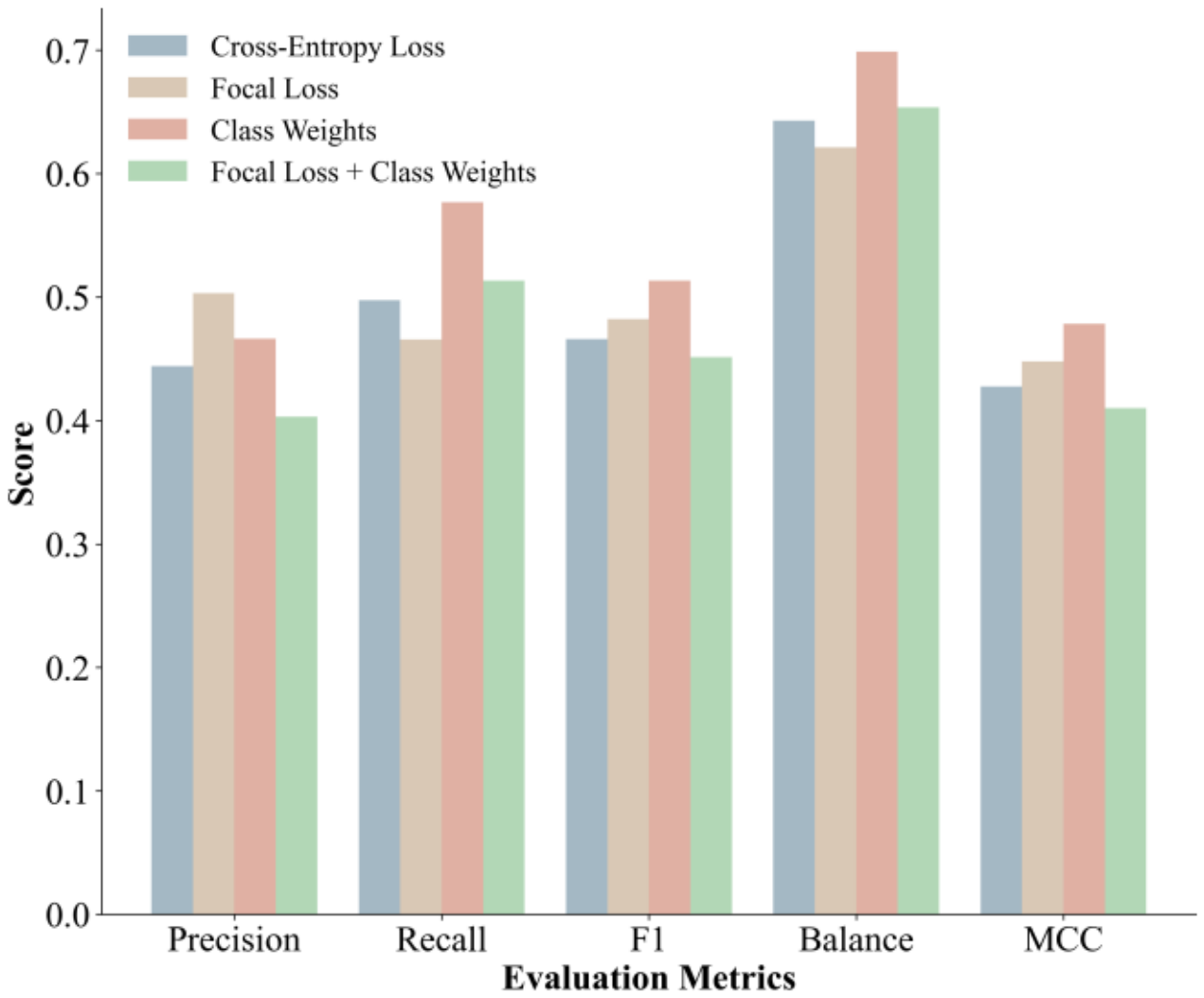

**Figure 3: Impact of More Advanced Techniques on Bioinformatics Software Defect Detection**

Additionally, combining Focal Loss and Class Weights led to unexpected performance degradation (see Figure 3). Our analysis suggests that excessive loss amplification for defective samples caused overfitting, increasing false positives and ultimately harming overall model performance.

**Findings-RQ2:** Compared to existing defect detection datasets, BioDefect is more effective for bioinformatics software defect detection and can significantly enhance performance.

# 5 DISSCUSSIONS AND THREATS TO VALIDITY

## 5.1 Discussions

This study represents the first exploration of defect detection in bioinformatics software and introduces the BioDefect dataset to address the limitations of existing datasets in this domain. Experimental results reveal a significant performance gap between existing datasets and BioDefect, highlighting the uniqueness of bioinformatics software, including differences in programming languages, coding conventions, and defect patterns. These factors hinder the effectiveness of existing datasets in defect detection for bioinformatics software. In contrast, BioDefect, which incorporates the characteristics of bioinformatics software, enables more precise identification of relevant defects.

The effectiveness of BioDefect demonstrates that constructing domain-specific datasets can significantly improve defect detection performance. Compared to general-purpose defect detection datasets such as Devign [12] and ReVeal [9], BioDefect focuses specifically on bioinformatics software and encapsulates domain-specific defect patterns, allowing models to better adapt to the complexities of this field. Even for C/C++ projects like BWA and Bowtie2, BioDefect still provides performance improvements, indicating that defect patterns in bioinformatics software differ significantly from those in general-purpose software. This suggests that relying solely on existing general datasets is insufficient for effective defect detection in bioinformatics software.

Additionally, while LLMs excel in code generation and understanding tasks, our experimental results indicate that simply increasing model parameters does not directly enhance defect detection performance in bioinformatics software. In this study, LLMs such as DeepSeek-R1 [10] and StarCoder2 [11] underperformed compared to smaller LMs like OPT [14] in defect detection. This could be attributed to two factors. First, LLMs are primarily trained to capture the overall semantics of code, whereas defect detection requires fine-grained recognition of error patterns. Second, smaller models rely more on fine-tuning data, making them better suited for adapting to domain-specific datasets like BioDefect. In contrast, due to differences in pre-training objectives, LLMs may not fully leverage such datasets. However, we observed that LLMs exhibit greater stability in cross-language datasets, suggesting their potential advantage in cross-language knowledge transfer, which warrants further investigation.

To optimize performance, we experimented with Focal Loss and Class Weights to increase the model's focus on defect samples. While these techniques led to some improvements, the gains were limited. This suggests that the complex code structures and sparse defect patterns in bioinformatics software make defect detection particularly challenging. Future research should explore novel strategies, such as designing model architectures better suited for bioinformatics software and integrating more domain-specific knowledge.

## 5.2 Threats to Validity

The defect category labeling in BioDefect has certain limitations. The accuracy of category labels relies on the subjective judgment of two domain experts, which introduces the possibility of bias or errors. Although previous studies [17] have shown that manual labeling is generally more accurate than automated methods, BioDefect may still contain a small number of mislabeled instances. Additionally, compared to some existing datasets, BioDefect is relatively small in scale, which may limit the generalization capability of models. In fact, we are currently conducting manual labeling for another large-scale bioinformatics software project. However, as this process is not yet complete, the dataset was not included in the experiments presented in this paper.

Another potential threat to validity is that our findings are based on a selected set of bioinformatics software, which may not generalize to all bioinformatics software. Different subfields within bioinformatics may have distinct coding styles and software development practices, which could influence defect patterns. Furthermore, although we evaluated multiple LMs, differences in model architectures or training configurations may affect the results. Future studies will expand the range of bioinformatics software examined and evaluate a broader set of model architectures to improve the generalizability of our findings.

# 6 RELATED WORK

**Defect Detection Datasets.** Existing datasets are primarily focused on C/C++ programs. For example, Devign [12] was constructed by Zhou et al., who hired four security researchers to manually annotate defect-related code from four C/C++ open-source projects. This process required 600 working hours and resulted in a dataset containing 58,965 function samples, with 27,652 labeled as defective, accounting for 46.9% of the total. Additionally, ReVeal [9] and Real-Vul [33] mainly focused on large-scale, general-purpose C/C++ projects, such as Linux and Chromium. The advantage of these datasets lies in their extensive coverage and real-world defect sources. However, they primarily target security vulnerabilities rather than general software defects, making them less applicable to defect detection in bioinformatics software due to the significant differences in defect patterns. In contrast, BioDefect is derived directly from real-world open-source bioinformatics software, aligning more closely with the unique characteristics of bioinformatics code and defect distributions. As a result, BioDefect is better suited to the task of bioinformatics software defect detection. The introduction of this dataset fills a gap in existing defect detection resources for bioinformatics software and provides a more targeted training resource for future research.

**Defect Detection Methods.** With the success of deep neural networks in image recognition and natural language processing, researchers have explored deep learning techniques for defect detection. Li et al. [43] proposed VulDeePecker, which employs deep learning to identify security vulnerabilities. Their approach detects unsafe function calls in code as entry points for analysis, extracts code slices, vectorizes them using word2vec [44], and then trains a Bi-LSTM model. Experimental results showed that VulDeePecker outperforms traditional methods by reducing false positive rates. In subsequent studies [45, 46], researchers tested various model architectures, including bi-GRU and DBN, and found that Bi-GRU achieved the best defect detection performance, identifying 15 vulnerabilities across four real-world projects. Inspired by image classification techniques, Wu et al. [47] proposed a novel defect detection method that effectively transforms function source code into an image format while preserving program details, improving accuracy and scalability in large-scale code defect detection. However, the complex contextual dependencies and long-distance relationships in bioinformatics code may limit the effectiveness of deep learning approaches in this domain.

More recently, researchers have attempted to leverage LMs and fine-tuning techniques to improve program analysis [48, 49]. Kanade et al. [49] introduced CuBERT, a pre-trained model accompanied by an open-source benchmark suite consisting of five classification tasks and one program repair task. Among them, three classification tasks focus on defect detection, including variable misuse classification, incorrect binary operator usage, and incorrect swapped operand detection. Feng et al. [36] proposed CodeBERT, a BERT-based pre-trained code model trained on both programming languages and natural languages. To evaluate the applicability of pre-trained models for non-training tasks, Zhou et al. [50] experimented with CodeBERT in defect detection tasks, demonstrating that simply replacing the encoder of an existing model with CodeBERT can achieve near state-of-the-art performance. This finding suggests that LMs can generalize beyond their original training tasks and be applied effectively to defect detection. Wang et al. [13] introduced CodeT5, a T5-based model designed to enhance code generation and understanding. They later extended this work with CodeT5+ [8], a more advanced version supporting multiple programming languages and various task modes. CodeT5+ has shown promising results on datasets like Devign. However, these approaches primarily focus on detecting whether a piece of code contains defects but do not precisely locate the specific lines where defects occur[51]. To address this limitation, An et al. [6] proposed a novel defect detection approach that integrates line-level defect localization into a unified training framework. By jointly learning defect detection and line-level defect localization, their model enhances knowledge sharing between tasks, achieving more fine-grained defect identification.

Although numerous studies have contributed to general software defect detection, to the best of our knowledge, no prior research has specifically explored defect detection for bioinformatics software. Our study fills this gap and provides a foundation for developing more effective defect detection techniques tailored to bioinformatics software.

# 7 CONCLUSIONS

This study represents the first exploration of defect detection in bioinformatics software and introduces BioDefect, the first dataset specifically designed for this task. Due to differences in programming languages, coding conventions, and defect patterns, existing datasets struggle to perform well in bioinformatics software defect detection. Experimental results demonstrate that BioDefect significantly enhances defect detection performance for bioinformatics software, addressing the limitations of general-purpose datasets in this domain. Furthermore, our findings emphasize that increasing model size does not necessarily lead to better performance. Instead, domain-specific datasets play a crucial role in improving defect detection capabilities.

# REFERENCES

[1] Teresa K Attwood, Sarah Blackford, Michelle D Brazas, Angela Davies, and Maria Victoria Schneider. 2019. A global perspective on evolving bioinformatics and data science training needs. Briefings in Bioinformatics 20, 2, 398-404. https://doi.org/10.1093/bib/bbx100

[2] Xu-Kai Ma, Yan Yu, Tao Huang, Dake Zhang, Caihuan Tian, Wenli Tang, Ming Luo, Pufeng Du, Guangchuang Yu, and Li Yang. 2024. Bioinformatics software development: Principles and future directions. The Innovation Life 2, 3, 100083. https://doi.org/10.59717/j.xinn-life.2024.100083

[3] Adeeb Noor. 2022. Improving bioinformatics software quality through incorporation of software engineering practices. PeerJ Computer Science 8e839. https://doi.org/10.7717/peerj-cs.839

[4] Christof Koch, and Allan Jones. 2016. Big Science, Team Science, and Open Science for Neuroscience. Neuron 92, 3, 612-616. https://doi.org/10.1016/j.neuron.2016.10.019

[5] Michael Fu, and Chakkrit Tantithamthavorn. 2022. Linevul: A transformer-based line-level vulnerability prediction. In Proceedings of the Proceedings of the 19th International Conference on Mining Software Repositories, 2022, 608-620.

[6] Jimin An, YunSeok Choi, and Jee-Hyong Lee. 2024. Code Defect Detection Using Pre-trained Language Models with Encoder-Decoder via Line-Level Defect Localization. In Proceedings of the Proceedings of the 2024 Joint International Conference on Computational Linguistics, Language Resources and Evaluation (LREC-COLING 2024), May, 2024, Torino, Italia. ELRA and ICCL, 3446-3456.

[7] Benjamin Steenhoek, Md Mahbubur Rahman, Shaila Sharmin, and Wei Le. 2023. Do Language Models Learn Semantics of Code? A Case Study in Vulnerability Detection. ArXiv abs/2311.04109

[8] Yue Wang, Hung Le, Akhilesh Gotmare, Nghi Bui, Junnan Li, and Steven Hoi. 2023. CodeT5+: Open Code Large Language Models for Code Understanding and Generation. In Proceedings of the Proceedings of the 2023 Conference on Empirical Methods in Natural Language Processing, December, 2023, Singapore. Association for Computational Linguistics, 1069-1088. https://doi.org/10.18653/v1/2023.emnlp-main.68

[9] Saikat Chakraborty, Rahul Krishna, Yangruibo Ding, and Baishakhi Ray. 2022. Deep Learning Based Vulnerability Detection: Are We There Yet? IEEE Transactions on Software Engineering 48, 9, 3280-3296. https://doi.org/10.1109/TSE.2021.3087402

[10] Daya Guo, Dejian Yang, Haowei Zhang, Junxiao Song, Ruoyu Zhang, Runxin Xu, Qihao Zhu, Shirong Ma, Peiyi Wang, and Xiao Bi. 2025. DeepSeek-R1: Incentivizing Reasoning Capability in LLMs via Reinforcement Learning. ArXiv abs/2501.12948

[11] Anton Lozhkov, Raymond Li, Loubna Ben Allal, Federico Cassano, Joel Lamy-Poirier, Nouamane Tazi, Ao Tang, Dmytro Pykhtar, Jiawei Liu, Yuxiang Wei, Tianyang Liu, Max Tian, Denis Kocetkov, Arthur Zucker, Younes Belkada, Zijian Wang, Qian Liu, Dmitry Abulkhanov, Indraneil Paul, Zhuang Li, Wen-Ding Li, Megan L. Risdal, Jia Li, Jian Zhu, Terry Yue Zhuo, Evgenii Zheltonozhskii, Nii Osae Osae Dade, W. Yu, Lucas Krauss, Naman Jain, Yixuan Su, Xuanli He, Manan Dey, Edoardo Abati, Yekun Chai, Niklas Muennighoff, Xiangru Tang, Muhtasham Oblokulov, Christopher Akiki, Marc Marone, Chenghao Mou, Mayank Mishra, Alexander Gu, Binyuan Hui, Tri Dao, Armel Randy Zebaze, Olivier Dehaene, Nicolas Patry, Canwen Xu, Julian J. McAuley, Han Hu, Torsten Scholak, Sébastien Paquet, Jennifer Robinson, Carolyn Jane Anderson, Nicolas Chapados, Mostofa Patwary, Nima Tajbakhsh, Yacine Jernite, Carlos Muñoz Ferrandis, Lingming Zhang, Sean Hughes, Thomas Wolf, Arjun Guha, Leandro von Werra, and Harm de Vries. 2024. StarCoder 2 and The Stack v2: The Next Generation. ArXiv abs/2402.19173

[12] Yaqin Zhou, Shangqing Liu, Jingkai Siow, Xiaoning Du, and Yang Liu. 2019. Devign: effective vulnerability identification by learning comprehensive program semantics via graph neural networks. In Proceedings of the Proceedings of the 33rd International Conference on Neural Information Processing Systems, 2019. Curran Associates Inc., 10197 - 10207.

[13] Yue Wang, Weishi Wang, Shafiq Joty, and Steven C.H. Hoi. 2021. CodeT5: Identifier-aware Unified Pre-trained Encoder-Decoder Models for Code Understanding and Generation. In Proceedings of the Proceedings of the 2021 Conference on Empirical Methods in Natural Language Processing, November, 2021, Online and Punta Cana, Dominican Republic. Association for Computational Linguistics, 8696-8708. https://doi.org/10.18653/v1/2021.emnlp-main.685

[14] Susan Zhang, Stephen Roller, Naman Goyal, Mikel Artetxe, Moya Chen, Shuohui Chen, Christopher Dewan, Mona T. Diab, Xian Li, Xi Victoria Lin, Todor Mihaylov, Myle Ott, Sam Shleifer, Kurt Shuster, Daniel Simig, Punit Singh Koura, Anjali Sridhar, Tianlu Wang, and Luke Zettlemoyer. 2022. OPT: Open Pre-trained Transformer Language Models. ArXiv abs/2205.01068

[15] T. Y. Lin, P. Goyal, R. Girshick, K. He, and P. Dollár. 2020. Focal Loss for Dense Object Detection. IEEE Transactions on Pattern Analysis and Machine Intelligence 42, 2, 318-327. https://doi.org/10.1109/TPAMI.2018.2858826

[16] Yizheng Chen, Zhoujie Ding, Lamya Alowain, Xinyun Chen, and David Wagner. 2023. DiverseVul: A New Vulnerable Source Code Dataset for Deep Learning Based Vulnerability Detection. In Proceedings of the Proceedings of the 26th International Symposium on Research in Attacks, Intrusions and Defenses, 2023, Hong Kong, China. Association for Computing Machinery, 654–668. https://doi.org/10.1145/3607199.3607242

[17] Yangruibo Ding, Yanjun Fu, Omniyyah Ibrahim, Chawin Sitawarin, Xinyun Chen, Basel Alomair, David Wagner, Baishakhi Ray, and Yizheng Chen. 2024. Vulnerability Detection with Code Language Models: How Far Are We? ArXiv abs/2403.18624

[18] Hao-Nan Zhu, and Cindy Rubio-González. 2023. On the Reproducibility of Software Defect Datasets. In Proceedings of the 2023 IEEE/ACM 45th International Conference on Software Engineering (ICSE), 14-20 May 2023, 2023, 2324-2335. https://doi.org/10.1109/ICSE48619.2023.00195

[19] René Just, Darioush Jalali, and Michael D. Ernst. 2014. Defects4J: a database of existing faults to enable controlled testing studies for Java programs. In Proceedings of the Proceedings of the 2014 International Symposium on Software Testing and Analysis, 2014, San Jose, CA, USA. Association for Computing Machinery, 437–440. https://doi.org/10.1145/2610384.2628055

[20] Ruchika Malhotra, Sonali Chawla, and Anjali Sharma. 2023. Software defect prediction using hybrid techniques: a systematic literature review. Soft Computing 27, 12, 8255-8288. https://doi.org/10.1007/s00500-022-07738-w

[21] Nima Shiri Harzevili, Alvine Boaye Belle, Junjie Wang, Song Wang, Zhen Ming Jiang, and Nachiappan Nagappan. 2024. A Systematic Literature Review on Automated Software Vulnerability Detection Using Machine Learning. ACM Comput. Surv. https://doi.org/10.1145/3699711

[22] Michael Pradel, and Koushik Sen. 2018. DeepBugs: a learning approach to name-based bug detection. Proceedings of the ACM on Programming Languages 2, OOPSLA, 1-25. https://doi.org/10.1145/3276517

[23] Miltiadis Allamanis, Henry Jackson-Flux, and Marc Brockschmidt. 2021. Self-supervised bug detection and repair. Advances in Neural Information Processing Systems 3427865-27876.

[24] Jiahao Fan, Yi Li, Shaohua Wang, and Tien N. Nguyen. 2020. A C/C++ Code Vulnerability Dataset with Code Changes and CVE Summaries. In Proceedings of the 2020 IEEE/ACM 17th International Conference on Mining Software Repositories (MSR), 25-26 May 2020, 2020, 508-512. https://doi.org/10.1145/3379597.3387501

[25] Miltiadis Allamanis. 2019. The adverse effects of code duplication in machine learning models of code. In Proceedings of the Proceedings of the 2019 ACM SIGPLAN International Symposium on New Ideas, New Paradigms, and Reflections on Programming and Software, 2019, Athens, Greece. Association for Computing Machinery, 143–153. https://doi.org/10.1145/3359591.3359735

[26] Roland Croft, M. Ali Babar, and M. Mehdi Kholoosi. 2023. Data Quality for Software Vulnerability Datasets. In Proceedings of the Proceedings of the 45th International Conference on Software Engineering, 2023, Melbourne, Victoria, Australia. IEEE Press, 121–133. https://doi.org/10.1109/icse48619.2023.00022

[27] Shaojie Yang, Haoran Xu, Fangliang Xu, and Yongjun Wang. 2024. S2Vul: Vulnerability Analysis Based on Self-supervised Information Integration. In Proceedings of the 2024 IEEE 35th International Symposium on Software Reliability Engineering (ISSRE), 28-31 Oct. 2024, 2024, 84-95. https://doi.org/10.1109/ISSRE62328.2024.00019

[28] F. Alexander Wolf, Philipp Angerer, and Fabian J. Theis. 2018. SCANPY: large-scale single-cell gene expression data analysis. Genome Biology 19, 1, 15. https://doi.org/10.1186/s13059-017-1382-0

[29] H. Li, and R. Durbin. 2009. Fast and accurate short read alignment with Burrows-Wheeler transform. Bioinformatics (Oxford, England) 25, 14, (2009/05/20), 1754-1760. https://doi.org/10.1093/bioinformatics/btp324

[30] Ben Langmead, and Steven L. Salzberg. 2012. Fast gapped-read alignment with Bowtie 2. Nature Methods 9, 4, 357-359. https://doi.org/10.1038/nmeth.1923

[31] Shuai Lu, Daya Guo, Shuo Ren, Junjie Huang, Alexey Svyatkovskiy, Ambrosio Blanco, Colin Clement, Dawn Drain, Daxin Jiang, and Duyu Tang. 2021. Codexglue: A machine learning benchmark dataset for code understanding and generation. ArXiv abs/2102.04664

[32] Benjamin Steenhoek, Md Mahbubur Rahman, Richard Jiles, and Wei Le. 2023. An Empirical Study of Deep Learning Models for Vulnerability Detection. In Proceedings of the Proceedings of the 45th International Conference on Software Engineering, 2023, Melbourne, Victoria, Australia. IEEE Press, 2237–2248. https://doi.org/10.1109/icse48619.2023.00188

[33] Partha Chakraborty, Krishna Kanth Arumugam, Mahmoud Alfadel, Meiyappan Nagappan, and Shane McIntosh. 2024. Revisiting the Performance of Deep Learning-Based Vulnerability Detection on Realistic Datasets. IEEE Transactions on Software Engineering 50, 8, 2163-2177. https://doi.org/10.1109/TSE.2024.3423712

[34] Daya Guo, Shuo Ren, Shuai Lu, Zhangyin Feng, Duyu Tang, Shujie Liu, Long Zhou, Nan Duan, Jian Yin, Daxin Jiang, and M. Zhou. 2020. GraphCodeBERT: Pre-training Code Representations with Data Flow. ArXiv abs/2009.08366

[35] Jacob Devlin, Ming-Wei Chang, Kenton Lee, and Kristina Toutanova. 2019. BERT: Pre-training of Deep Bidirectional Transformers for Language Understanding. In Proceedings of the Proceedings of the 2019 Conference of the North American Chapter of the Association for Computational Linguistics:

Human Language Technologies, Volume 1 (Long and Short Papers), June, 2019, Minneapolis, Minnesota. Association for Computational Linguistics, 4171-4186. https://doi.org/10.18653/v1/N19-1423

[36] Zhangyin Feng, Daya Guo, Duyu Tang, Nan Duan, Xiaocheng Feng, Ming Gong, Linjun Shou, Bing Qin, Ting Liu, Daxin Jiang, and Ming Zhou. 2020. CodeBERT: A Pre-Trained Model for Programming and Natural Languages. In Proceedings of the Findings of the Association for Computational Linguistics: EMNLP 2020, 2020, Online. Association for Computational Linguistics, 1536–1547. https://doi.org/10.18653/v1/2020.findings-emnlp.139

[37] Colin Raffel, Noam Shazeer, Adam Roberts, Katherine Lee, Sharan Narang, Michael Matena, Yanqi Zhou, Wei Li, and Peter J. Liu. 2020. Exploring the limits of transfer learning with a unified text-to-text transformer. The Journal of Machine Learning Research 21, 1, Article 140.

[38] Erik Nijkamp, Bo Pang, Hiroaki Hayashi, Lifu Tu, Haiquan Wang, Yingbo Zhou, Silvio Savarese, and Caiming Xiong. 2023. CodeGen: An Open Large Language Model for Code with Multi-Turn Program Synthesis. In Proceedings of the International Conference on Learning Representations, 2023,

[39] Thomas Wolf, Lysandre Debut, Victor Sanh, Julien Chaumond, Clement Delangue, Anthony Moi, Pierric Cistac, Tim Rault, Remi Louf, Morgan Funtowicz, Joe Davison, Sam Shleifer, Patrick von Platen, Clara Ma, Yacine Jernite, Julien Plu, Canwen Xu, Teven Le Scao, Sylvain Gugger, Mariama Drame, Quentin Lhoest, and Alexander Rush. 2020. Transformers: State-of-the-Art Natural Language Processing. In Proceedings of the Proceedings of the 2020 Conference on Empirical Methods in Natural Language Processing: System Demonstrations, October, 2020, Online. Association for Computational Linguistics, 38-45. https://doi.org/10.18653/v1/2020.emnlp-demos.6

[40] Davide Chicco, Matthijs J. Warrens, and Giuseppe Jurman. 2021. The Matthews Correlation Coefficient (MCC) is More Informative Than Cohen's Kappa and Brier Score in Binary Classification Assessment. IEEE Access 978368-78381. https://doi.org/10.1109/ACCESS.2021.3084050

[41] Shiqi Tang, Song Huang, Changyou Zheng, Erhu Liu, Cheng Zong, and Yixian Ding. 2022. A novel cross-project software defect prediction algorithm based on transfer learning. Tsinghua Science and Technology 27, 1, 41-57. https://doi.org/10.26599/TST.2020.9010040

[42] Bo Li, Yongqiang Yao, Jingru Tan, Gang Zhang, Fengwei Yu, Jianwei Lu, and Ye Luo. 2022. Equalized focal loss for dense long-tailed object detection. In Proceedings of the Proceedings of the IEEE/CVF conference on computer vision and pattern recognition, 2022, 6990-6999.

[43] Zhen Li, Deqing Zou, Shouhuai Xu, Xinyu Ou, Hai Jin, Sujuan Wang, Zhijun Deng, and Yuyi Zhong. 2018. VulDeePecker: A Deep Learning-Based System for Vulnerability Detection. In Proceedings of the Proceedings 2018 Network and Distributed System Security Symposium, 2018, https://doi.org/10.14722/ndss.2018.23165

[44] Tomas Mikolov, Kai Chen, Gregory S. Corrado, and Jeffrey Dean. 2013. Efficient Estimation of Word Representations in Vector Space. In Proceedings of the International Conference on Learning Representations, 2013,

[45] Z. Li, D. Zou, S. Xu, Z. Chen, Y. Zhu, and H. Jin. 2022. VulDeeLocator: A Deep Learning-Based Fine-Grained Vulnerability Detector. IEEE Transactions on Dependable and Secure Computing 19, 4, 2821-2837. https://doi.org/10.1109/TDSC.2021.3076142

[46] Z. Li, D. Zou, S. Xu, H. Jin, Y. Zhu, and Z. Chen. 2022. SySeVR: A Framework for Using Deep Learning to Detect Software Vulnerabilities. IEEE Transactions on Dependable and Secure Computing 19, 4, 2244-2258. https://doi.org/10.1109/TDSC.2021.3051525

[47] Y. Wu, D. Zou, S. Dou, W. Yang, D. Xu, and H. Jin. 2022. VulCNN: An Image-inspired Scalable Vulnerability Detection System. In Proceedings of the 2022 IEEE/ACM 44th International Conference on Software Engineering (ICSE), 25-27 May 2022, 2022, 2365-2376. https://doi.org/10.1145/3510003.3510229

[48] Daya Guo, Shuai Lu, Nan Duan, Yanlin Wang, Ming Zhou, and Jian Yin. 2022. UniXcoder: Unified Cross-Modal Pre-training for Code Representation. In Proceedings of the Proceedings of the 60th Annual Meeting of the Association for Computational Linguistics (Volume 1: Long Papers), May, 2022, Dublin, Ireland. Association for Computational Linguistics, 7212-7225. https://doi.org/10.18653/v1/2022.acl-long.499

[49] Aditya Kanade, Petros Maniatis, Gogul Balakrishnan, and Kensen Shi. 2020. Learning and evaluating contextual embedding of source code. In Proceedings of the Proceedings of the 37th International Conference on Machine Learning, 2020. JMLR.org, 5110-5121.

[50] Xin Zhou, DongGyun Han, and David Lo. 2021. Assessing Generalizability of CodeBERT. In Proceedings of the 2021 IEEE International Conference on Software Maintenance and Evolution (ICSME), 27 Sept.-1 Oct. 2021, 2021, 425-436. https://doi.org/10.1109/ICSME52107.2021.00044

[51] Fangcheng Qiu, Zhongxin Liu, Xing Hu, Xin Xia, Gang Chen, and Xinyu Wang. 2024. Vulnerability Detection via Multiple-Graph-Based Code Representation. IEEE Transactions on Software Engineering 50, 8, 2178-2199. https://doi.org/10.1109/TSE.2024.3427815